\documentclass[fleqn,12pt,twoside]{article}
\usepackage{espcrc1}
\usepackage{fleqn,espcrc1}
\usepackage{graphicx}
\usepackage[figuresright]{rotating}
\usepackage{epsfig}
\newcommand{\vgas}{\ensuremath{\mathrm V_{gas}}}
\newcommand{\qgamma}{\ensuremath{\mathrm Q_\gamma}}

\title{Preliminary results of an aging test of RPC chambers for the
  LHCb Muon System}

\author{A. Bizzeti\address[FIRE]{\small University of Florence and
  INFN -- Florence, Via G. Sansone 1, I--50019, Sesto F.no, Florence,
  Italy}, G. Carboni\address[ROM2]{\small University of Rome ``Tor Vergata''
  and INFN -- Rome II, Via della Ricerca Scientifica 1, I--00133, Rome, 
  Italy}, G. Collazuol\addressmark[FIRE], S. De
  Capua\addressmark[ROM2], D. Domenici\addressmark[ROM2],
  G. Ganis\addressmark[ROM2]
  \thanks{Corresponding author: fax:~+39062025364, tel:~+390672594537, 
                                e-mail:~gerardo.ganis@cern.ch}, 
R. Messi\addressmark[ROM2], G. Passaleva\addressmark[FIRE], E. Santovetti\addressmark[ROM2], M. Veltri\addressmark[FIRE] %
}%
      
\begin{document}

\maketitle

\begin{abstract}
{\small 
{\bf Abstract.} The preliminary results of an aging test performed at the
CERN Gamma Irradiation Facility on a single--gap RPC prototype
developed for the LHCb Muon System are presented. The results are based on an
accumulated charge density of 0.42 C/cm$^2$, corresponding to about 4
years of LHCb running at the highest background rate. 
We observe a rise in the dark current and noise 
measured with source off. 
The current drawn with source on steadily decreased, possibly indicating 
an increase of resistivity of the chamber plates.
The performance of the chamber, studied with a muon beam under 
several photon flux values, 
is found to still fulfill the LHCb operation requirements.

}
\vspace{1pc}
\end{abstract}

{\bf Introduction. }
The LHCb experiment~\cite{TP} has chosen to equip
about 48\% of the Muon System~\cite{TDR} with  
Resistive Plate Chamber (RPC) detectors.
These detectors will be subject to a very large particle flux, 
potentially dangerous from the point of view of detector aging 
\footnote{
Aging in RPCs is due mostly to the current 
flowing through the resistive plates; 
it has been shown, indeed, that the
irradiation of bakelite slabs with photons up to an integrated dose of
20 kGy does not produce any degradation in the bakelite
properties~\cite{danilo}.}. 
In the regions
covered by RPCs, the maximum particle rate is expected to vary
between 0.19 in region~4 and 0.75 kHz/cm$^2$
in region~3 (see~\cite{TDR} for a definition of the regions). 
This leads, 
under reasonable hypothesis
\footnote{ Namely: {\it i)} running at half maximum rate; 
{\it ii)} assuming an average avalanche charge of 30 pC~\cite{danilo}.},
to total integrated charge densities between 
0.35 C/cm$^2$ (region~4) and 1.1 C/cm$^2$ (region 4) for 10 
years running. These charge densities are  
significantly larger than those expected by
ATLAS~\cite{atlas_tdr} and CMS experiments~\cite{cms_tdr}, whose 
aging tests accumulated at most 
about 0.3 C/cm$^2$~\cite{ATLASCMSaging}, clearly 
insufficient to draw conclusions about LHCb RPCs. 
An extensive aging program has been therefore devised for the LHCb needs 
exploiting the large CERN Gamma Irradiation Facility (GIF)~\cite{GIF}, where a
$^{137}$Cs gamma source of about 655 GBq is available.
This paper reports the preliminary results of the first 
part of the test performed with a prototype chamber during 
January-December 2001. The second part, currently under preparation, 
will be done with close-to-final  
LHCb chambers and is expected to start March 2002. 

\begin{figure}
\begin{picture}(1,1)
\put(10,250){ {\it a)}}
\put(10, 80){ {\it c)}}
\put(10,-90){ {\it e)}}
\put(230,250){ {\it b)}}
\put(230, 80){ {\it d)}}
\put(230,-90){ {\it f)}}
\put(350,-105){ {\scriptsize HV=10200 V}}
\put(350,-160){ {\scriptsize HV=10400 V}}
\put(350,-215){ {\scriptsize HV=10600 V}}
\end{picture}
\scalebox{1.}{
\begin{tabular}{cc}
 \includegraphics[width=5.0cm,height=5.0cm]{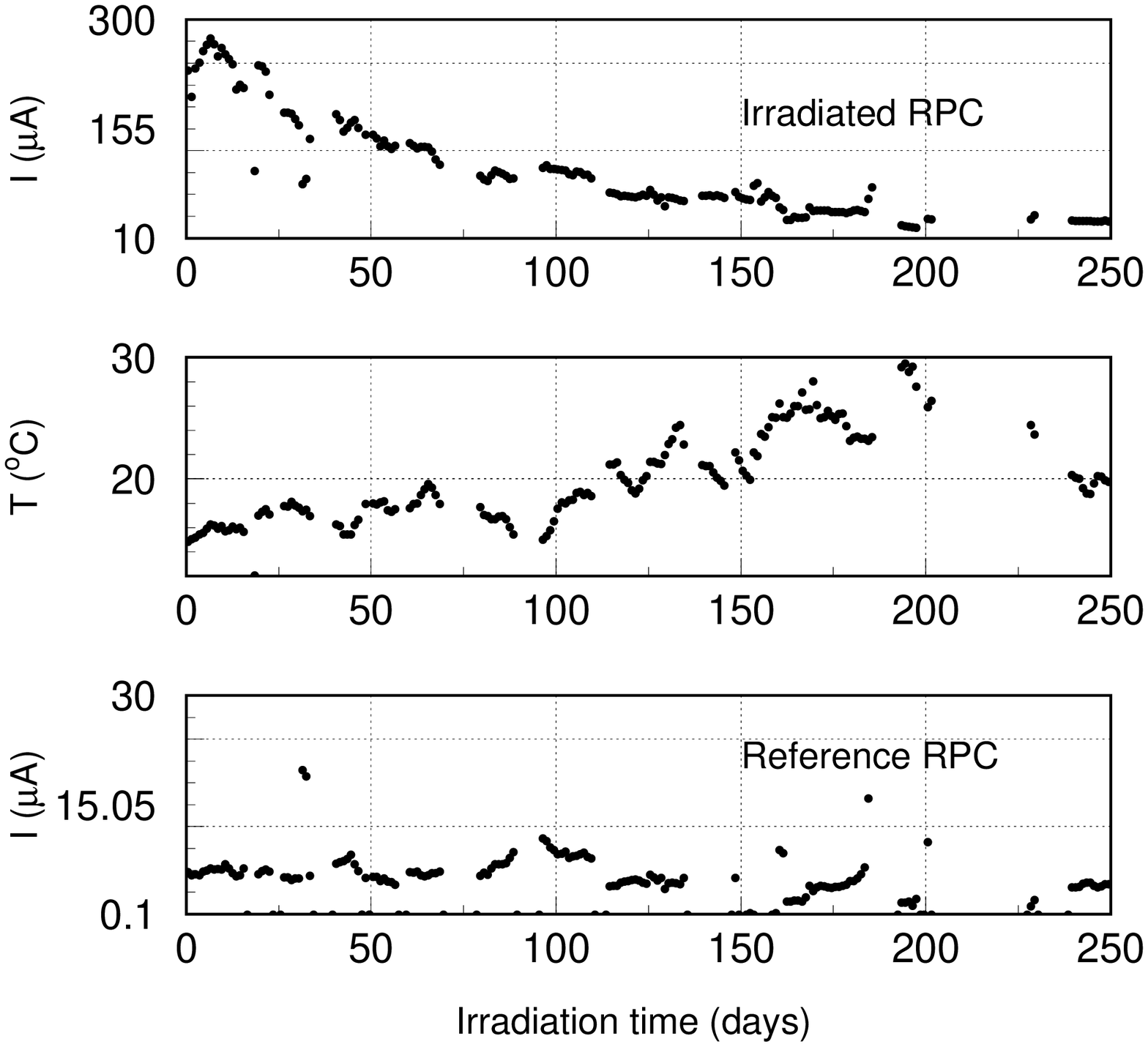} &
 \includegraphics[width=5.0cm,height=5.0cm]{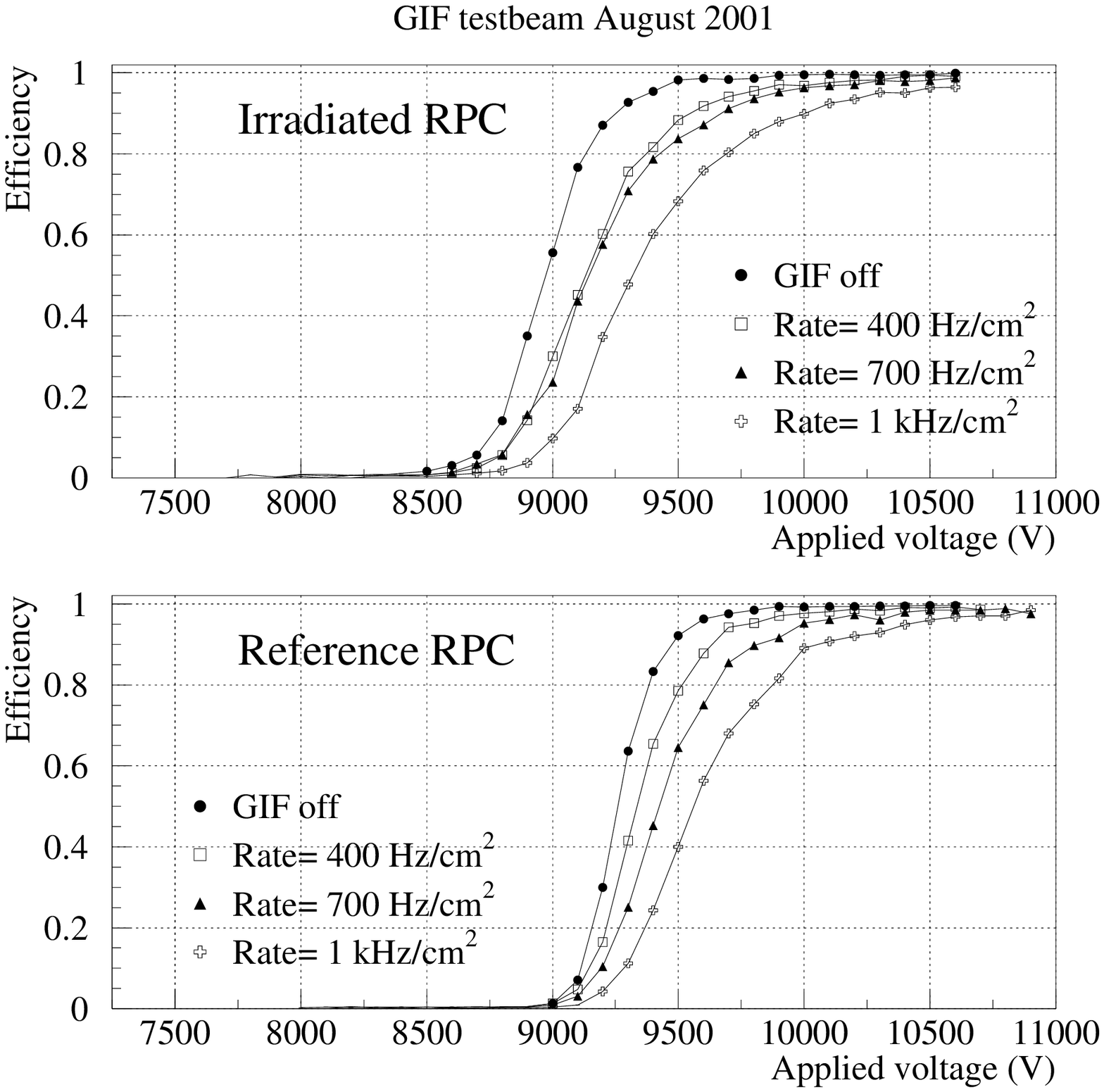} \\
 \includegraphics[width=5.0cm,height=5.0cm]{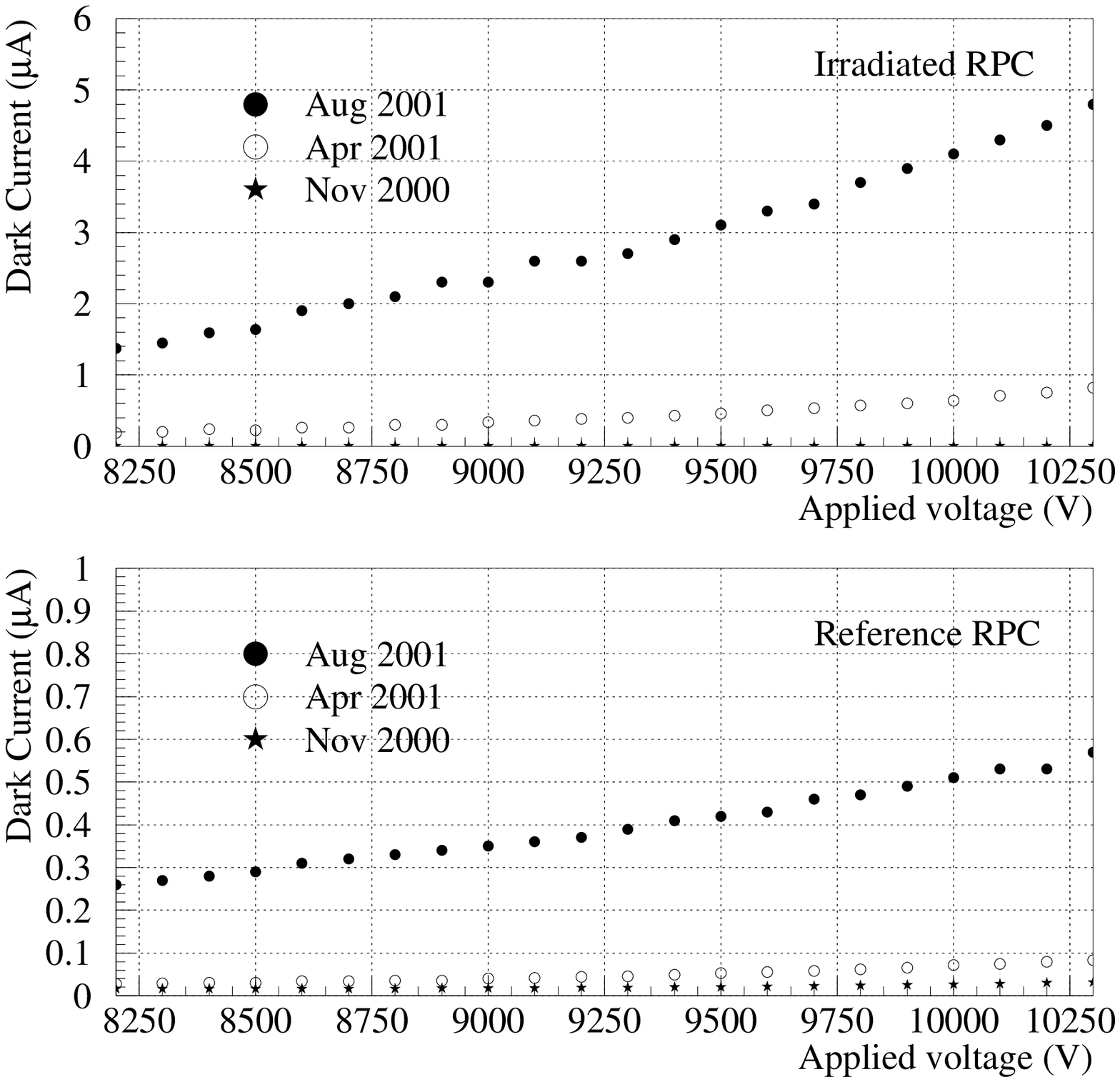} &
 \includegraphics[width=5.0cm,height=5.0cm]{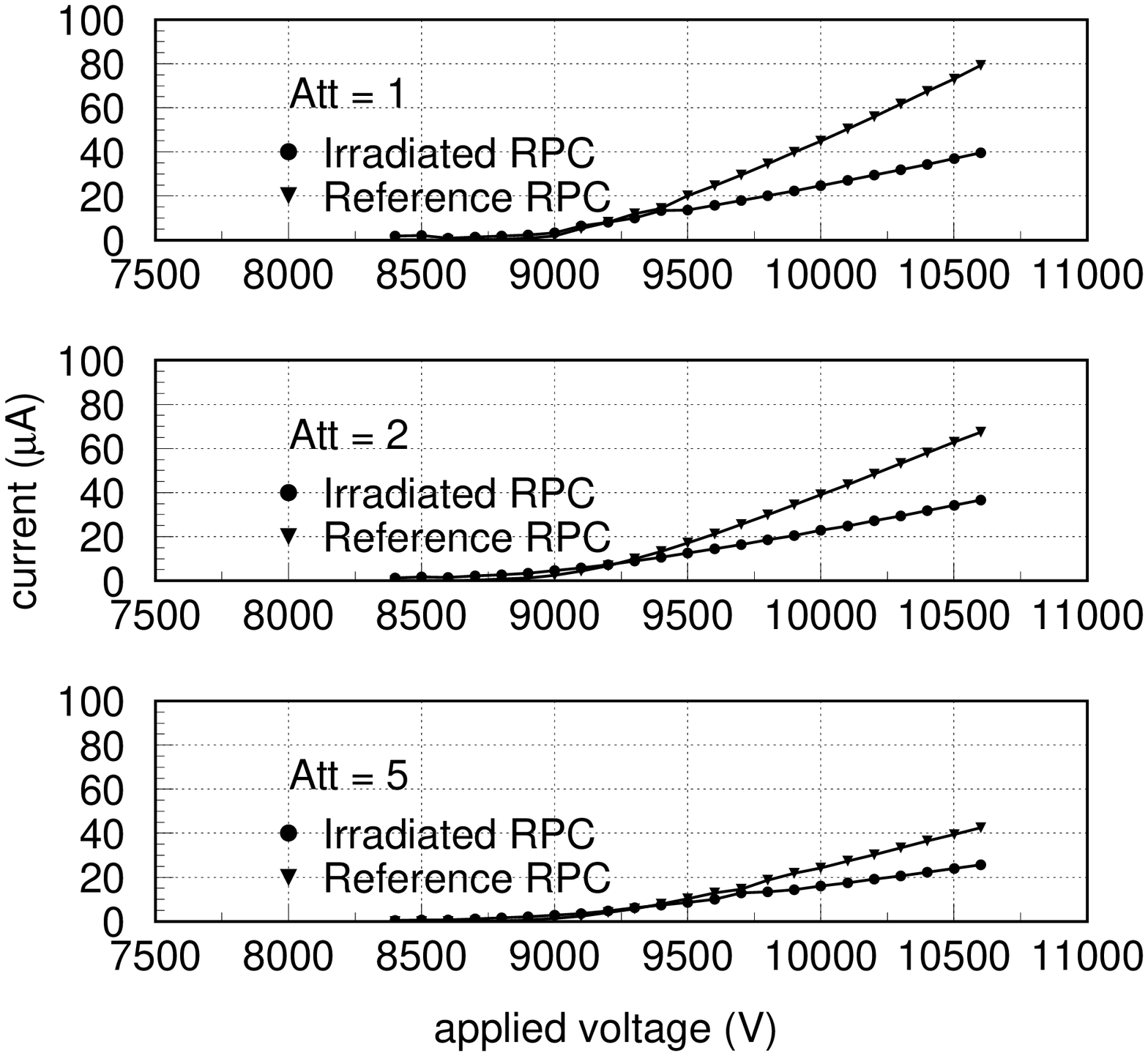} \\
 \includegraphics[width=5.0cm,height=5.0cm]{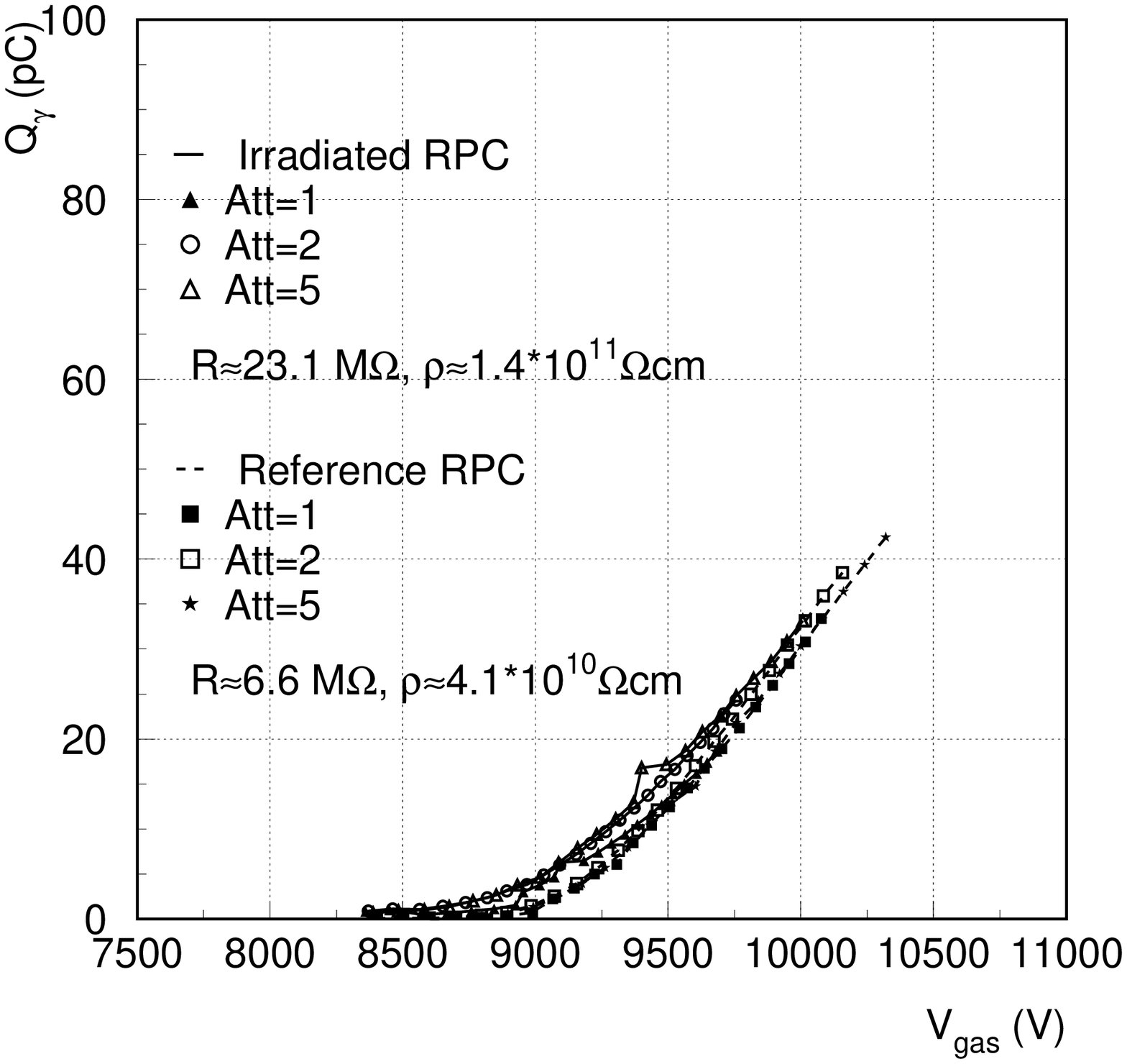} &
 \includegraphics[width=5.0cm,height=5.0cm]{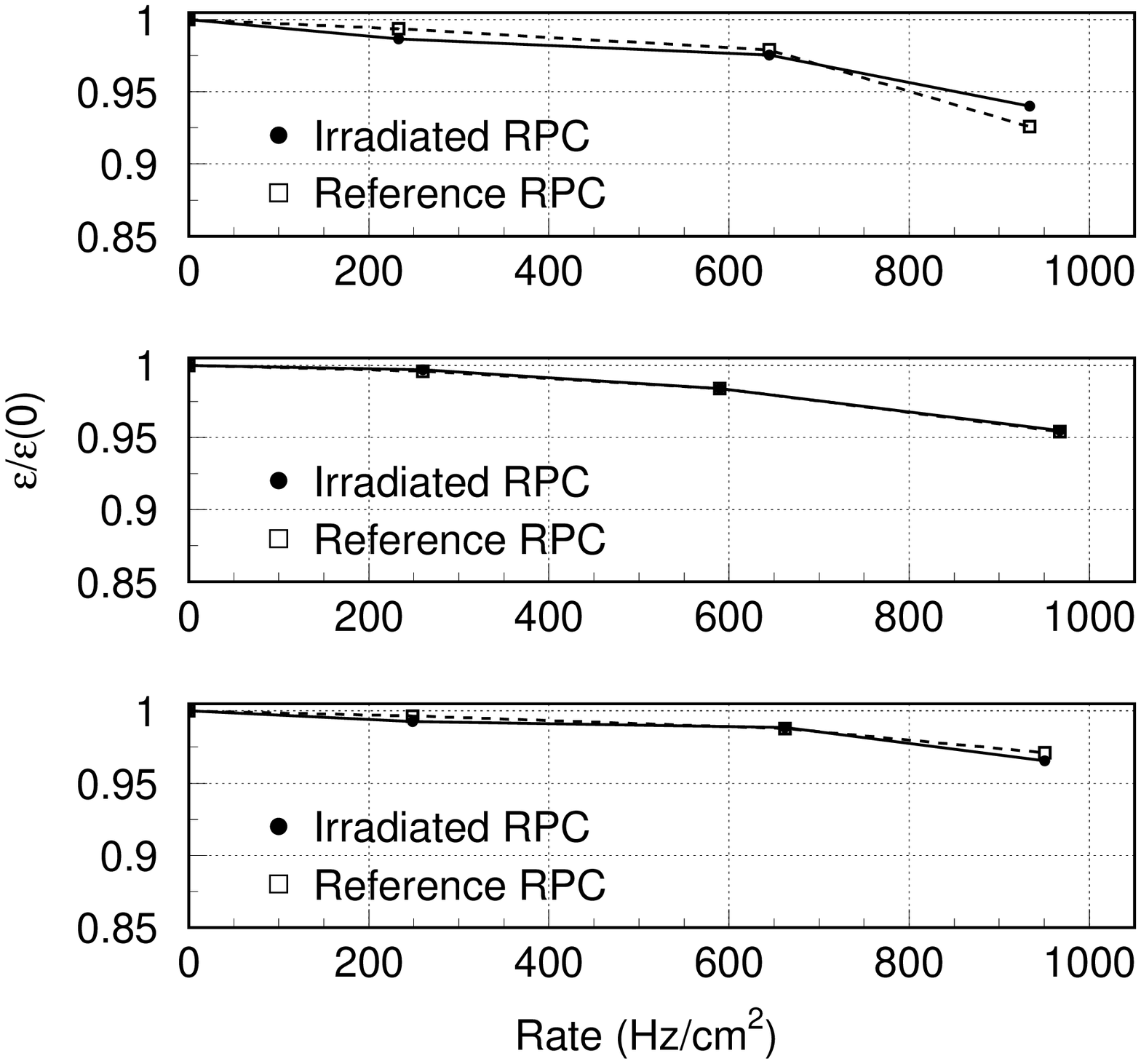}
\end{tabular}
}
\caption{{\small Panel {\it a)}: relevant quantities monitored during the aging test;
  {\it top}: current drawn by the irradiated RPC  
   as a function of time; {\it middle}: temperature; 
   {\it bottom}: current drawn by the reference chamber. 
   Panel {\it b)}: chamber efficiencies measured at GIF at 
    the end of the aging test under several photon fluxes. 
   Panel {\it c)}: dark currents, measured before (Nov 2000), 
   during (Apr 2001) and at the end (Aug 2001) of the aging test. 
  Panel {\it d)}: RPC currents under source irradiation at different
   photon fluxes measured at the end of the aging test. 
  Panel {\it e)}: average avalanche charge \qgamma\ as a function of \vgas\  
   measured at the end of the aging test under several photon fluxes. 
  Panel {\it f)}: efficiency versus rate for the two RPCs at three different
   high voltage values; the efficiencies are normalized to those
   obtained with the source off.}
}
 \label{fig:one} 
\end{figure}

{\bf Setup of the aging test. }
The prototype RPC used for the test ({\it irradiated RPC}) 
is a single gap oiled chamber, with active area 50x50 cm$^2$, made 
with 2~mm thick phenolic bakelite plates with 
nominal resistivity $\rho = 9\cdot 10^9\
\Omega$~cm. This RPC was placed in front of the source at a distance 
varying between 0.5-1~m and a photon flux of about 1-2 kHz/cm$^2$.  
To facilitate disentangling the effects due to variations in the environment
parameters - such as temperature, pressure or gas quality - from those due
to the irradiation, a second chamber ({\it reference RPC}), 
identical to the irradiated one, 
was placed just outside the irradiation area. 
The RPCs were operated with a gas mixture consisting of 95\% $\rm
C_2H_2F_4$, 4\% $\rm i\!-\!C_4H_{10}$, 1\% $\rm SF_6$.
To simulate the LHCb conditions and perform the test in a reasonable
amount of time, the applied voltage was chosen in such a way to
have an average avalanche charge $\qgamma$ of about 50 pC, 
which, at the rate quoted above,
yields a current density of about 80 nA/cm$^2$. 
%
%
All the relevant parameters of the test - 
temperature, pressure, high voltage and currents drawn by the chambers - 
were continuously recorded. In
Fig.~\ref{fig:one}a the current drawn by the 
irradiated RPC~ and by the reference RPC during the aging test are 
shown as a function of time; 
these currents have been corrected for temperature variations, 
assuming I=I$_0$~exp($\alpha$T) with $\alpha$ chosen such that 
intra-day fluctuations are eliminated. 
As it is clearly visible from the top plot, the current drawn by the 
irradiated RPC decreased with time. 
The current drawn
by the reference RPC is reasonably
stable and constant during the test.
By integrating the measured current,   
the total charge accumulated by the irradiated RPC has been 
found to be 0.42 C/cm$^2$, representing  
about 4 LHCb years in regions~3
and to more than 10 LHCb years in regions~4.
The charge integrated by the reference RPC is 0.02 C/cm$^2$. 

{\bf Measurements after the aging test. }
The general behavior and performance of the irradiated RPC 
has been checked using the source and the muon beam available 
at the GIF.
The combined use of the
gamma source and the beam allows a
test of the chamber performance 
under background conditions very similar to the ones encountered in the experiment. 
Measurements were performed with 
different photon fluxes, ranging from zero
(source off) up to a maximum of about 1
kHz/cm$^2$. The chambers were operated in the same conditions as in the aging
test.
The efficiency was evaluated by tracking the beam particles
with a scintillator hodoscope and with an additional $10\times
10$ cm$^2$ RPC. The photon fluxes were estimated from the chamber
counting rates measured during dedicated off-spill gates.
Data were collected at three
different source attenuation values, namely attenuation 1, 2 and 5
corresponding roughly to 1 kHz/cm$^2$, 0.7 kHz/cm$^2$ and 0.4
kHz/cm$^2$ photon fluxes respectively.

The performances of the irradiated RPC are still very good and 
basically indistinguishable from the ones of reference chamber; 
as an example 
the efficiency curves are shown in Fig.~\ref{fig:one}b for different 
photon fluxes 
\footnote{The shift in the plateau position was observed also before 
the irradiation test and is probably due to a relative mis-calibration of the two 
sets of readout electronics; it is also compatible with a difference in 
gap size within tolerances.}. 

In Fig.~\ref{fig:one}c the dark currents drawn by the two RPCs
as a function of the applied voltage - measured before, during and at the end 
of the aging test - are shown. A large increase
of the dark current drawn by the irradiated RPC can be clearly
observed. The effect of the irradiation amounts, at a possible working point,
 to about 
1.5 nA/cm$^2$. This increase is correlated with a corresponding 
increase in dark noise rate, but its origin is still under investigation.
However, this increase in dark current density is still significantly below 
the limit of 3 nA/cm$^2$ set in the LHCb-$\mu$ TDR~\cite{TDR}.

The currents drawn by the two chambers under irradiation and 
different photon fluxes 
are shown in Fig.~\ref{fig:one}d: the irradiated RPC 
clearly draws less current than the reference one, 
indicating that \qgamma{}, which can be 
obtained from the ratio of current to measured-hit-rate, 
is smaller in the irradiated RPC.   

The effect can be interpreted within a model introduced in
Ref.~\cite{aielli1} in which it is assumed 
that, under a high particle flux, the RPC working point is
determined by an effective voltage $\rm \vgas = V - IR $
where V is the applied voltage, I the current drawn by the RPC
and R the resistance of the resistive plates. 
Under these assumptions, quantities such as \qgamma\ are universal 
functions of the gas properties and of \vgas. 
Figure~\ref{fig:one}e shows
that this assumption works reasonably well when the appropriate value 
for R is used: $\rm R_{irr} = 23.1\, M\Omega$ and
$\rm R_{ref} = 6.6\, M\Omega$ respectively corresponding to equivalent
bakelite resistivities $\rm \rho_{irr} = 1.4\times 10^{11}\, \Omega cm$
and $\rm \rho_{ref} =  4.1\times 10^{10}\, \Omega cm$.
This model, therefore, suggests that the radiation caused a significant 
increase of resistance in the irradiated RPC. 
%
%
%
%
%

An increase in resistivity would result in a limitation of the 
rate capability of the detectors. 
Figure~\ref{fig:one}f shows the efficiency of the 
two chambers, normalized to the value obtained with source off, 
as a function of the photon flux for three different high voltage values. 
No evidence of a different
behavior between the two RPCs up to a flux of
more than 1 kHz/cm$^2$ is observed, showing that 
these RPC can satisfy the  
LHCb requirements at least for 4 years running 
in the worst expected background conditions.

{\bf Summary and conclusions. }
A charge density of 0.42 C/cm$^2$ has been accumulated on a single gap, 
oiled, low-resistivity RPC.  
Dark currents and noise rates, measured with source off, clearly rose. 
The current measured with source on steadily decreased, possibly indicating 
an increase of the bakelite resistivity. 
The performances of the chamber are not affected at least up to 
1 kHz/cm$^2$, still satisfying the LHCb requirements. 
The integrated charge corresponds to
about 4 years of LHCb at the highest expected background rate. 

\vspace{.5cm}
{\it Acknowledgments.} 
We would like to thank V.~Souvorov and T.~Schneider for valuable help 
and suggestions during preparation and running of the test.

\newpage

\vspace{-20cm}

\end{document}